\documentclass[aps,prl]{revtex4}
\usepackage{amssymb}

\usepackage{amsmath}
\usepackage{graphicx}
\usepackage{epsfig,amsmath}

\begin{document}

\title{Evolution of a Network of Vortex Loops in HeII. Exact Solution of the ''Rate
Equation''.}
\author{Sergey K. Nemirovskii}
\date{today}

\begin{abstract}
Evolution of a network of vortex loops in HeII due to the fusion
and breakdown of vortex loops is studied. We perform investigation
on the base of \ the ''rate equation'' for the distribution
function $n(l)$ of number of loops of length $l$ proposed by
Copeland with coauthors. By using the special ansatz in the
''collision'' integral we have found the exact power-like solution
of ''kinetic equation'' in stationary case. That solution is the
famous equilibrium distribution $n(l)\varpropto l^{-5/2}$ obtained
earlier in numerical calculations. Our result, however, is not
equilibrium, but on the contrary, it describes the state with two
mutual fluxes of the length (or energy) in space of the vortex
loop sizes. Analyzing this solution we drew several results on
the structure and dynamics of the vortex tangle in the superfluid
turbulent helium. In particular, we obtained that the mean radius
of the curvature is of order of interline space. We also obtain
that the decay of the vortex tangle obeys the Vinen equation,
obtained earlier phenomenologically. We evaluate also the full
rate of reconnection events. \newline PACS-number 67.40
\end{abstract}

\maketitle

\affiliation{Institute for Thermophysics, Lavrentyeva, 1, 630090
Novosibirsk, Russia}

A network of one-dimensional singularities appears in various physical
systems affecting important properties of them. As examples we would point
out quantized vortices in quantum fluids both in turbulent regime (see e.g.
book \cite{Donbook}and papers \cite{Feynman}, \cite{Vinen}, \cite{NF}) and
thermodynamically equilbrium\ state (\cite{Zurek96}, \cite{Nemir05}). Other
examples are the flux tubes in superconductors \cite{Kleinert}, dislocations
in solids \cite{Nabarro67}, global cosmic strings \cite{Copeland98},\cite
{Sreer99} polymer chains \cite{Wiegel73}. To evaluate contribution of
mentioned structure in various effects, such as thermodynamic and kinetic
properties, phase transition etc. we have to know structure and dynamics of
network of singularities. \newline
The whole evolution of network of chaotic set of line consists of two main ingredients.
The first is the motion of the individual elements of each of lines, due to
specific equations of motion, different for each of cases listed above. For
instance elements of vortex filaments in turbulent superfluid $^{4}$He move
obeying the Biot-Savart law supplemented by the friction force and the
external flow/counterflow if any. Besides of motion of elements there
is another very important element of the whole dynamics, common for all systems -
fusion and breakdown of the loops or recombination of open strings due to
reconnection processes. Further for definiteness we will talk about vortex
loops in superfluid helium.\newline
It is widely appreciated that the reconnection processes influence both the
structure and evolution of vortex tangle. For instance, Feynman in his
pioneering paper devoted to superfluid turbulence proposed scenario how the
vortex tangle decays in absence of the external counterflow. According to
this scenario a fusion of small vortex rings into larger ones as well as a
breakdown into smaller ones is possible at the moments of the reconnection
events. In assumption that \textit{\ on the average} the last property
dominates, the cascade like process of formation of smaller and smaller
loops forms. When the scale of the small rings becomes of the order of the
interatomic distances, which is the final stage of the cascade, the vortex
motion is degenerated into thermal excitations. The idea that degeneration
of the vortex tangle occurs due to cascade-like transferring of the length
in space of scale of sizes of vortex loops was indirectly confirmed only in
numerical calculations, where the procedure of artificial elimination of
small loops had been used \cite{Schwarz88}-\cite{Tsubota00}.\newline
\qquad In spite of the recognized importance of the reconnecting loops
kinetics, the numerical results remain main source of information about this
process. The obvious lack of theoretical investigations interferes with the
deep insight in the nature of this phenomena (this situation had been
recently discussed in \cite{Barenghi2004}). For instance it is not clear how
the cascade of length in space of vortex loops sizes is formed, what
mechanisms are responsible for this, what quantities determine  intensity
of cascade, and why at all the breakdown of the loop prevail. The scarcity of
analytic investigations related to incredible complexity of the problem.
Indeed we have to deal with a set of objects with not fixed number of
elements, they can born and die. Thus, some analog of the secondary
quantization method is required with the difference that objects (vortex
loops) themselves possess an infinite number of degree of freedom with very
involved dynamics. Clearly this problem can hardly be resolved in nearest
future. Recently in \cite{Copeland98} much more modest approach, based on
the ''rate equation'' for distribution function $n(l)$ was elaborated in
context of cosmic strings. Following this work we introduce distribution
function $n(l,t)$ of density of loop in ''space'' of their lengths. It is
defined as the number of loops (per unit volume) with lengths lying between
$l$ and $l+dl$. Due to reconnection processes $n(l,t)$ can vary. We
discriminate two types of processes, namely the fusion of two loops into the
larger single loop and the breakdown of single loop into two daughter loops.
The kinetic of vortex tangle is affected by the intensity of the introduced
processes. The intensity of the first process is characterized by the rate
of collision $A(l_{1},l_{2},l)$ of two loops with lengths $l_{1}$ and $l_{2}$
and forming the loop of length $\ l=l_{1}+l_{2}$. The intensity of the
second process is characterized by the rate of self-intersection $%
B(l,l_{1},l_{2})$ of the loop with length $l$ into two daughter loops with
lengths $\ l_{1}$ and $l_{2}$. In view of exposed above we can directly
write out the master ''kinetic'' equation for rate of change the
distribution function $n(l,t)$.
\begin{gather}
\frac{\partial n(l,t)}{\partial t}=\int \int
A(l_{1},l_{2},l)n(l_{1})n(l_{2})\delta
(l-l_{1}-l_{2})dl_{1}dl_{2}\;\;\;\;\;\;\;\;\;\;\;\ \ \ \ \
\;l_{1}+l_{2}\rightarrow l\;\;\ \ \ \ \ \ \;(1)  \notag \\
-\int \int A(l_{1},l,l_{2},)\delta
(l_{2}-l_{1}-l)n(l)n(l_{1})dl_{1}dl_{2}\;\;\;\;\;\;\;\;\;\;\ \ \ \ \ \ \ \ \
\;\ \ \ \ \ \ \;\;l_{1}+l\rightarrow l_{2}\;\;\;\;\ \ \ \ \;(2)
\label{Master} \\
-\int \int A(l_{2},l,l_{1},)\delta
(l_{1}-l_{2}-l)n(l)n(l_{1})dl_{1}dl_{2}\;\;\;\;\;\;\;\;\;\;\;\;\ \ \ \ \ \ \
\ \ \ \ \ \ \ \ \ l_{2}+l\rightarrow l_{1}\;\;\;\;\;\ \ \ (3)  \notag \\
-\int \int B(l_{1},l_{2},l)n(l)\delta
(l-l_{1}-l_{2})dl_{1}dl_{2}\;\;\;\;\;\;\;\;\;\ \ \ \;\;\;\ \;\;\;\;\;\;\;\;\
\ \ \ \ \ \ \ \ \;l\rightarrow l_{1}+l_{2}\;\;\;\ \ \ \ \ \ \;(4)  \notag \\
+\int \int B(l,l_{2},l_{1})\delta
(l_{1}-l-l_{2})n(l_{1})dl_{1}dl_{2}\;\;\;\;\;\;\;\ \;\;\;\;\;\;\;\;\;\;\ \ \
\ \ \ \ \ \ \ \;l_{1}\rightarrow l+l_{2}\;\;\;\;\;\;\ \ \ \ \ \ (5)  \notag
\\
+\int \int B(l,l_{1},l_{2})\delta
(l_{2}-l-l_{1})n(l_{1})dl_{1}dl_{2}\;.\;\;\;\
\;\;\;\;\;\;\;\;\;\;\;\;\;\;\;\;\;\;\;\;\ \ \ \ \ \ \ \ \ \ \ \ \ \
\;l_{2}\rightarrow l+l_{1}\;\;\;\;\;\;\;\;\;(6)\;\;\;\;\;\;\;  \notag
\end{gather}
\ All processes are depicted at the left of each line. Coefficient $A$
and $B$ \ were evaluated on the base of qualitative considerations in \cite
{Copeland98}. A bit more rigorous way, having universal character was
elaborated in work \cite{Nemir_05_RR}. Both models give similar expressions
of type
\begin{equation}
A(l_{1},l_{2},l)=b_{m}V_{l}l_{1}l_{2},\;\ \ B(l_{1},l-l_{1},l)=b_{s}\frac{%
V_{l}l}{(\xi _{0}l_{1})^{3/2}}.  \label{A_and_B}
\end{equation}
Here $V_{l}$ is some characteristic velocity ov elements of line,
\ $b_{m}$ and $b_{s}$ are some constants depending on the model.
For instance in work \cite{Copeland98} \ on evolution of network
of cosmic strings autors offered values about $0.1\div 0.3$ for
both coefficients (not necessary equal). In paper
\cite{Nemir_05_RR} on vortex loops in superfluid turbulent HeII
there was offered $b_{m}\sim 1/3,$ $b_{s}\approx 0.0164772.$ The
quantity $\xi _{0} $ appears in as the persistency length of
theory of randomly walking chains modelling chaotic cosmic
strings. For the vortex tangle in superfluid helium this
quantity \ is associated with the mean radius of curvature (see
\cite{Nemirovskii_97_1}). Both these approaches are failed for scales near $%
\xi _{0},$ therefore usually this value appears as a low cut-off. \ Equation
(\ref{Master}) \ had been studied (mainly numerically) in papers \cite
{Copeland98}, \cite{Sreer99}, where \ some conclusions about evolution of
network of cosmic string were made.\newline
\qquad In the present work we demonstrate that the master ''rate equation'' (%
\ref{Master}) has exact stationary power-like solution of form
$n(l)=C\ast l^{s}$. We discuss the physical meaning of this
solution and apply it to describe some properties of vortex tangle
in the superfluid turbulent helium. To find power-like solution of
form $n(l)=C\ast l^{s}$ we use the Zakharov ansats, which \ is the
special treatment of the ''collision'' integral in equation
(\ref{Master}). This trick was elaborated by Zakharov for the wave
turbulence (see e.g. \cite{Zakharov_book}), now we will show how
it works in our case. Let us take for instance first and second
integrals in the ''collision term'' of (\ref{Master}). Let us
further perform in the second integral the following change of
variables.
\begin{equation}
l=\tilde{l}_{2}\left( \frac{l}{\tilde{l}_{2}}\right) ,\;\ \ \ l_{1}=\tilde{l}%
_{1}\left( \frac{l}{\tilde{l}_{2}}\right) ,\;\;\ \;l_{2}=l\left( \frac{l}{%
\tilde{l}_{2}}\right) .  \label{ansatz}
\end{equation}
Under this change of variables various factors in the integrand of the
second integral transforms as follows
\begin{equation*}
\delta (l_{2}-l_{1}-l)\rightarrow \left( \frac{l}{\tilde{l}_{2}}\right)
^{-1}\delta (l-\tilde{l}_{1}-\tilde{l}_{2}).
\end{equation*}
\begin{equation*}
n(l)\rightarrow n(\tilde{l}_{2})\left( \frac{l}{\tilde{l}_{2}}\right)
^{s},\;\;\;\;\;n(l_{1})\rightarrow n(\tilde{l}_{1})\left( \frac{l}{\tilde{l}%
_{2}}\right) ^{s},
\end{equation*}
\begin{equation*}
A(l_{1},l,l_{2})\rightarrow \frac{1}{2}V_{l}\tilde{l}_{1}\tilde{l}_{2}\left(
\frac{l}{\tilde{l}_{2}}\right) ^{2}=A(\tilde{l}_{1},\tilde{l}_{2},l)\left(
\frac{l}{\tilde{l}_{2}}\right) ^{2}.
\end{equation*}
As result the second integral in the ''collision'' term takes a form
(additional term $3$ in the power counting appears from the Jacobian of
transformation)
\begin{equation}
\int \int \left( \frac{l}{\tilde{l}_{2}}\right) ^{2+2s-1+3}A(\tilde{l}_{1},%
\tilde{l}_{2},l)n(\tilde{l}_{1})n(\tilde{l}_{2})\delta (l-\tilde{l}_{1}-%
\tilde{l}_{2})d\tilde{l}_{1}d\tilde{l}_{2}\;.\;  \label{ansatz_2}
\end{equation}
It is easy to see that the transformed second term \bigskip\ in the
''collision'' integral in rhs of master kinetic equation (\ref{Master})
turns into first integral with additional factor $\left( l/\tilde{l}%
_{2}\right) ^{4+2s}$ in the integrand. Performing the same
procedure for all lines we conclude that the ''collision integral'' of
the ''rate equation'' can be written as
\begin{eqnarray}
&&\int \int A(l_{1},l_{2},l)n(l_{1})n(l_{2})\left( 1-\left( \frac{l}{l_{1}}%
\right) ^{4+2s}-\left( \frac{l}{l_{2}}\right) ^{4+2s}\right) \delta
(l-l_{1}-l_{2})dl_{1}dl_{2}\;\;\;\;\;  \label{kin_power} \\
&&-\int \int B(l_{1},l_{2},l)n(l)\left( 1-\left( \frac{l}{l_{1}}\right)
^{s+3/2}-\left( \frac{l}{l_{2}}\right) ^{s+3/2}\right) \delta
(l-l_{1}-l_{2})dl_{1}dl_{2}\;\;\;\;\;\;\;\;\;\;\;\;\ \;\;  \notag
\end{eqnarray}
For $s=-5/2$ both expressions $1-\left( \frac{l}{l_{1}}\right)
^{4+2s}-\left( \frac{l}{l_{2}}\right) ^{4+2s}$ and $1-\left( \frac{l}{l_{1}}%
\right) ^{s+3/2}-\left( \frac{l}{l_{2}}\right) ^{s+3/2}$ are equal to $%
(l-l_{1}-l_{2})/l$. Thus the integrands of both integrals in (\ref{kin_power}%
) include expressions of type $\left( x\right) \delta (x)$ and
these integrals vanish. This implies in stationary case the power-like
solution $n=C\ast l^{-5/2}$ for distribution function $n(l,t)$ of
density of loop in ''space'' of their lengths takes place.\newline
Let us discuss the physical meaning of the solution obtained.
First of all we stress that it does not relate to detailed
balance, it rather corresponds to nonequilibrium state. To clarify what is
happening we introduce density $L(t) $ (in space of sizes $l$) of
full length per unit of volume
\begin{equation}
L(l,t)=l\ast n(l,t)=\frac{\text{total length}}{\text{unit of volume*interval
of length}}.  \label{dL/dl}
\end{equation}
The total length (per unit volume), or the vortex line density
\begin{equation*}
\mathcal{L}(t)=\int L(l,t)dl=\int l\ast n(l,t)dl
\end{equation*}
is obviously conserved during the reconnections events $\;d\mathcal{L}%
(t)/dt=0$. Conservation of the vortex line density can be
expressed in form of continuity equation:
\begin{equation}
\frac{\partial L(t)}{\partial t}+\frac{\partial P(l)}{\partial l}=0.
\label{continuity equation}
\end{equation}
This form of equation states that the rate of change of length is associated
with ''flux'' of length in space of sizes of the loops. Term ''flux'' here
means just redistribution of length among the loops due to reconnections.
Expression for $P(l)$ is obtained by multiplication kinetic equation \ by $l$
and by rewriting the ''collision'' term in the shape of derivative with
respect to $l$. \ Result is (substitutions $l_{1}/l=x$ and $l_{2}/l=y$ \ had
been used below)\newline
\begin{eqnarray}
&&P=\left( \frac{l^{5+2s}}{5+2s}\right) \int \int \frac{1}{2}%
b_{m}V_{l}xyC^{2}x^{s}y^{s}\left( 1-\left( \frac{1}{x}\right) ^{4+2s}-\left(
\frac{1}{y}\right) ^{4+2s}\right) \delta (1-x-y)dxdy  \label{flux_integral}
\\
&&-\left( \frac{l^{s+5/2}}{s+5/2}\right) \int \int \frac{1}{2}b_{s}V_{l}%
\frac{1}{(\xi _{0}x)^{3/2}}C\left( 1-\left( \frac{1}{x}\right)
^{s+3/2}-\left( \frac{1}{y}\right) ^{s+3/2}\right) \delta (1-x-y)dxdy.
\notag
\end{eqnarray}
Both integrals in relation (\ref{flux_integral}) coincide with integrals in (%
\ref{kin_power}), therefore they vanish for $\ s=-5/2$. However
they have pre-integral factor with the denominators, which also
vanish for $\ s=-5/2$. Calculating indeterminacy $0/0$ we obtain
the final expressions for ''flux'' of length in space of sizes of
the loops
\begin{equation}
P=\frac{12.555}{2}C^{2}b_{m}V_{l}-\frac{5.545}{2\xi _{0}{}^{3/2}}Cb_{s}V_{l}
\label{Flux}
\end{equation}
The positive sign of the first corresponds to flux of length in
direction of large scales. This is justified, since the fusion
processes lead to formation of large loops. The negative sign of
the second term corresponds to flux of length in direction of
small scales. This is justified, since the breaking down processes
lead to formation of small loops.\newline \qquad The approach
elaborated above allows to draw several conclusions concerning the
structure and dynamics of real vortex tangle in turbulent HeII. To
do it we have to specify quantity $V_{l}$ , which enters into the
rates coefficients of the both merging and breaking down
processes. Since the averaged radius of curvature is $\xi _{0}$,
we estimate the velocity factor $V_{l}$ to be of order of $\kappa
/\xi _{0}$ \ ($\kappa $ is the quantum of circulation). Thus the
only parameters of the whole theory are the quantum of circulation
$\kappa $\ and the mean radius of curvature $\xi _{0}$.\newline
\qquad {\large Vortex Line Density. }In steady case the positive
flux of length in (\ref{Flux}) exactly compensates the negative
flux. Equating these two terms we have
\begin{equation}
C=\frac{5.455}{12.555}\frac{b_{s}}{b_{m}}\frac{1}{\xi _{0}^{3/2}}=C_{VLD}%
\frac{1}{\xi _{0}^{3/2}}.  \label{C_constant}
\end{equation}
New numerical parameter $C_{VLD}\approx 0.2\div 1$ . Thus the power -like
solution of the master Kinetic equation $n(l)$ is
\begin{equation}
n(l)=\frac{C_{VLD}}{\xi _{0}^{3/2}}l^{-5/2}  \label{n(l)}
\end{equation}
Accordingly the total length \ $\mathcal{L}$ per unit of volume (we recall
that quantity $\xi _{0}$ serves as the low cut-off ).
\begin{equation}
\mathcal{L=}\int_{\xi _{0}}^{\infty }l\ast n(l)dl=\frac{2}{3}\frac{C_{VLD}}{%
\xi _{0}^{2}}.  \label{VLD_vs_curvature}
\end{equation}
Result (\ref{VLD_vs_curvature}) is remarkable. The idea that interline space
$\delta =\mathcal{L}^{-1/2}$ is of order of mean radius of curvature $\xi
_{0}$ was launched by Schwarz \cite{Schwarz88}. Earlier it was confirmed
only in numerical simulations and the nature of this phenomenon was not
clear. We proved that this relation appears due to kinetics of colliding
vortex loops.\newline
\qquad {\large Decay of Vortex Tangle (Vinen Equation). }Let us suppose that
there is some sink for loops of small sizes. Then the steady
situation is reached when there is some source which generates length. This
is the mutual friction for the superfluid turbulence. If we switch
off the source of \ length, the total length will decrease. The attenuation
of vortex line density is related to negative flux $P_{neg}$ (in direction
of small scales) of the full flux (\ref{Flux}) and can be evaluated from
the continuity equation for  density of length $\ L(l,t)$ in space
of the loop sizes, which we integrate over $l$\newline
\begin{equation*}
\frac{\partial L(l,t)}{\partial t}+\frac{\partial P(l)}{\partial l}%
=0\Rightarrow \frac{d\mathcal{L}(t)}{dt}+P_{neg}=0.
\end{equation*}
Evaluating $P_{neg}$ from relation for full flux \ (\ref{Flux}), where
constant \ $C$ is used from (\ref{C_constant}) we conclude that
\begin{equation}
\;\;\;\frac{d\mathcal{L}(t)}{dt}=-C_{Vinen}\kappa \mathcal{L}^{2}\;.\;
\label{VE}
\end{equation}
Here $C_{Vinen}$ is new numerical factor. For parameters, which we
adopt here $C_{Vinen}\approx 0.05\div 0.2$\bigskip. Expression
(\ref{VE}) is the famous Vinen equation \cite{Vinen} obtained him
from experiment. It is remarkable that this relation appears due
to the reconnection processes. \ The own dynamics of filament
specific for various systems is absorbed by the persistency length
$\xi _{0},$ which has dropped out the Vinen equation at all. Thus
the (\ref{VE}) has universal character and can be applied for
other systems, e.g. for cosmic strings. We also would like to note
that our calculations confirmed the brilliant Feynman's conjecture
on the formation of cascade-like breakdown of vortex
loops.\newline
\qquad {\large The Full Rate of Reconnection.}The full rate of reconnection $%
\dot{N}_{rec}$ can be evaluated directly from master ''rate equation'' (\ref
{Master}). Indeed, this equation describes change of $n(l)$ due to
reconnection events. It takes into account sign of events, depending on
whether the loop of size $l$ appears or dies in result of reconnection.
Therefore, if we take all terms in collision integral with the plus sign we
obtain the total number of reconnections. The according calculations lead to
result
\begin{equation*}
\dot{N}_{rec}=\frac{1}{3}\frac{\kappa (b_{s}C_{VLD}+b_{m}^{2}C_{VLD})}{\xi
_{0}^{5}}=C_{rec}\kappa \mathcal{L}^{5/2},
\end{equation*}
where $C_{rec}$ one more constant of order $0.1-0.5.$ That results agrees
with the recent numerical investigation by Barenghi and Samuels \cite
{Barenghi2004}.\newline
\qquad The problem of nonequilibrium dynamics of vortex loops, which merge
and break down due to reconnections has been considered. The description was
performed on the base of the ''rate equation'' for distribution of number of
loops in space of their lengths. By use of special substitution of variables
in the collision integral (Zakharov ansatz) we found the power-like solution
of the kinetic equation. That is non-equilibrium solution characterized by
two mutual fluxes of length (energy) in space of loop sizes.\newline
\qquad The result obtained were used to draw some conclusion about the
structure of the vortex tangle appeared in the superfluid turbulent HeII.(i)
In particular we obtained  that vortex line density is of order of inverse
squared mean curvature of lines. (ii) We also found that the rate of decay
of the vortex line density is proportional to squared vortex line density
itself multiplied by quantum of circulation and numerical factor of order of
unity. (iii) we also estimate the total number of reconnections, which \
turned out to be of \ order of quantum of circulation multiplied by the
vortex line density in power 5/2. \newline
\qquad I am grateful to participants of the workshop ''Superfluidity under
Rotation'' (Manchester, 2005) for useful discussion of the results exposed
above. This work was partially supported by grant N 03-02-16179 of the
Russian Foundation of Fundamental Research.


\begin{thebibliography}{99}
\bibitem{Donbook}  R.J.Donnelly, \textit{Quantized Vortices in HeliumII},
(Cambridge University Press, 1991).

\bibitem{Feynman}  R.P.Feynman, in \textit{Progress in Low Temperature
Physics}, edited by C.J.Gorter (North-Holland, Ameterdam, 1955), Vol.I, p.17.

\bibitem{Vinen}  W.F.Vinen, Proc. R. Soc. London A \textbf{242}, 493 (1957).

\bibitem{NF}  S.K.Nemirovskii and W.Fiszdon, Rev. Mod. Phys. \textbf{67}, 37
(1995).

\bibitem{Zurek96}  W. H. Zurek, Nature (London) \textbf{317}, 505 (1985);
Acta Phys. Pol. B \textbf{24}, 1301 (1993); Phys. Rep., \textbf{276},177,
(1996).

\bibitem{Nemir05}  S.K.Nemirovskii, Theoretical and Mathematical Physics,
\textbf{141,} 141, (2004).

\bibitem{Kleinert}  H. Kleinert, \textit{Gauge Fields in Condenced Matter
Physics}, {World Scientific Publishing, Singapore,1990}.

\bibitem{Nabarro67}  F.R.N. Nabarro,\textit{Theory of Crystal Dislocation},
(Oxford Univ. Press. Oxford, 1967)

\bibitem{Wiegel73}  F.W.Wiegel, \textit{Introduction to Path-Intrgral
Methods in Physics and Polymer Sciencs}, {World Scientific Publishing Co Pte
Ltd, 1986}.

\bibitem{Copeland98}  E.J.Copeland, T.W.B.Kibble and D.A.Steer, Phys.Rev. D,
textbf\{58, \} 043508, (1998).

\bibitem{Sreer99}  Joao Magueijoa, Havard Sandvik, and Dani`ele A. Steer,
http://lanl.arxiv.org/abs/astro-ph/9905363

\bibitem{Koplik}  J.Koplik and H.Levine, Phys. Rev. Lett., \textbf{71},
1375(1993).

\bibitem{Schwarz88}  K.W.Schwarz, Phys. Rev. B, \textbf{38}, 2398(1988).

\bibitem{Barenghi}  C.F.Barenghi, D.C.Samuels, G.H.Bauer and
R.J.Donnelly,Phys. Fluids \{ bf 9\}, 2631(1997).

\bibitem{Aarts}  R.G.K.M.Aarts and A.T.A.M.de Waele, Phys. Rev. B \textbf{50}%
,10069,(1994).

\bibitem{Tsubota00}  M. Tsubota, T. Araki and S. K. Nemirovskii, Phys. Rev.B
\textbf{62}, 11751 (2000).

\bibitem{Barenghi2004}  C.F.Barenghi and D.C.Samuels, Journal of Low
Temperature Physics, \textbf{136}, Nos. 5/6, September 2004

\bibitem{Nemir_05_RR}  S.K.Nemirovskii,
http://lanl.arxiv.org/abs/cond-mat/0505441.

\bibitem{Nemirovskii_97_1}  S.K. Nemirovskii, Phys. Rev \textbf{B57}, 5792
(1997).

\bibitem{Zakharov_book}  V.E. Zakharov, V.S. L'vov, G. Falkovich, \ \textit{%
Kolmogorov Spectra of Turbulence I}, Springer-Verlag, 1992
\end{thebibliography}
\end{document}